\documentclass[aps,prb,twocolumn,showpacs,superscriptaddress,floatfix]{revtex4-1}  
\usepackage[english]{babel}
\usepackage[T1]{fontenc}
\usepackage[utf8]{inputenc}
\usepackage{SIunits}
\usepackage{graphicx}
\usepackage{dcolumn}
\usepackage{amsmath}
\usepackage{natbib}
\usepackage{bm}
\usepackage{color}
\usepackage{textcomp}

\begin{document}

\title{A micro-magneto-Raman scattering study of graphene on a bulk graphite substrate}

\author{C. Faugeras}
\email{clement.faugeras@lncmi.cnrs.fr} \affiliation{LNCMI-CNRS
(UJF, UPS, INSA), BP 166, 38042 Grenoble Cedex 9, France}

\author{J. Binder}
\affiliation{LNCMI-CNRS (UJF, UPS, INSA), BP 166, 38042 Grenoble
Cedex 9, France}
\affiliation{Institute of Experimental Physics, Faculty of
Physics, University of Warsaw, Poland.}

\author{A. A. L. Nicolet}
\affiliation{LNCMI-CNRS (UJF, UPS, INSA), BP 166, 38042 Grenoble
Cedex 9, France}

\author{P. Leszczynski}
\affiliation{LNCMI-CNRS (UJF, UPS, INSA), BP 166, 38042 Grenoble
Cedex 9, France}

\author{P. Kossacki}
\affiliation{LNCMI-CNRS (UJF, UPS, INSA), BP 166, 38042 Grenoble
Cedex 9, France}
\affiliation{Institute of Experimental Physics, Faculty of
Physics, University of Warsaw, Poland.}

\author{A. Wysmolek}
\affiliation{Institute of Experimental Physics, Faculty of
Physics, University of Warsaw, Poland.}

\author{M. Orlita}
\affiliation{LNCMI-CNRS (UJF, UPS, INSA), BP 166, 38042 Grenoble
Cedex 9, France}

\author{M. Potemski}
\affiliation{LNCMI-CNRS (UJF, UPS, INSA), BP 166, 38042 Grenoble
Cedex 9, France}

\begin{abstract}

We report on a magneto-Raman scattering study of
graphene flakes located on the surface of a bulk graphite
substrate. By spatially mapping the Raman scattering response of
the surface of bulk graphite with an applied magnetic field, we
pinpoint specific locations which show the electronic excitation
spectrum of graphene. We present the characteristic Raman scattering
signatures of these specific locations. We show that such flakes can be superimposed with
another flake and still exhibit a graphene-like excitation spectrum.
 Two different excitation laser energies (514.5 and
720 nm) are used to investigate the excitation wavelength
dependence of the electronic Raman scattering signal.

\end{abstract}
\pacs{73.22.Lp, 63.20.Kd, 78.30.Na, 78.67.-n}

\maketitle

Soon after the first isolation of graphene and the experimental
demonstration of the integer quantum Hall effect of massless Dirac
fermions in this material~\cite{Novoselov2005,Zhang2005}, it was
realized that the substrate on which graphene is deposited has a
strong influence on the measured properties, in terms of
electronic mobility and of doping homogeneity~\cite{Martin2008}.
Different approaches have been used to increase the quality of
graphene, either by removing completely the substrate to create
suspended structures~\cite{Bolotin2008,Berciaud2009,Ki2013} or by
searching for a substrate more appropriate than SiO$_2$, such as
BN~\cite{Dean2010}. These two approaches were breakthroughs in the
research on graphene, allowing for the observation of the
fractional quantum Hall effect~\cite{Bolotin2009,Du2009,Dean2011}
and opening the area of studies of interacting Dirac fermions.
However, probably the most appropriate substrate for graphene is bulk
graphite, the 3D allotrope of sp$^2$ carbon. It is now known that
graphene flakes can be located on the surface of bulk graphite.
They have been identified by low temperature STM~\cite{Li2009}.
Infrared magneto-absorption experiments revealed a very low value
for the Fermi energy ($\sim 6$~meV) and an exceptionally high
electronic quality. These conclusions are based on the
experimental observation of Landau quantization at magnetic fields
as low as 1~mT and a cyclotron resonance absorption line width of
a few tenths of \textmu eV~\cite{Neugebauer2009} corresponding to an
electronic mobility above $10^7$ cm$^2$/(V.s). These two
characteristics have triggered a growing interest for this
graphene-like system.

More information on this system have been obtained with
magneto-Raman scattering experiments. Such techniques offer a
spatial resolution of $\sim 1$ \textmu m, well below the typical
size of the flakes. These experiments confirmed the graphene-like
excitation spectrum with the observation of the characteristic
sequence of magneto-phonon resonance as well as the direct
detection of purely electronic inter Landau level
excitations~\cite{Yan2010,Faugeras2011,Kuhne2012,Qiu2013}.
Nevertheless, some effects, unexpected in theory in pristine
graphene, have been observed such as the coupling of optical
phonons with $\Delta |n| =0$ inter Landau level
excitations~\cite{Faugeras2011}, where $n$ is the Landau level
index, with an apparent coupling strength that decreases with increasing
magnetic field, or the observation of $\Delta|n|=\pm 1$
excitations crossing the phonon feature without showing any signs
of an interaction~\cite{Faugeras2011,Kuhne2012}. All these
experiments called for further investigations of this system.

In this letter, we present a detailed Raman scattering study of graphene
flakes on a bulk graphite substrate. By spatially mapping the
surface of bulk graphite with an applied magnetic field, we
identify different criteria to locate such flakes and we describe
their electronic properties. We find that the Fermi velocity
characterizing the band structure of this graphene system can vary
by a few percents from one flake to another. Such flakes can
sit on top of each other and still give rise to a graphene-like
excitation spectrum. Based on experiments performed with different
excitation laser energies, we experimentally determine the
excitation wavelength evolution of the relative intensity of
phonon and electronic excitations and we compare it with
existing theoretical models.

The samples investigated are pieces of natural graphite which were
slightly exfoliated to remove the upper layers possibly exposed to
contamination. The sample is placed in a closed set-up with 200
mbars of He exchange gas at $T=4.2$~K. The Raman scattering
response is measured at the \textmu m scale with a home made Raman
scattering set-up based on a 5~\textmu m core mono-mode fiber connected to the
excitation laser (Ar$^+$ laser at 514.5 nm or Ti:Sapphire tuned to
720 nm), optical lenses, band pass filters and a dichroic mirror.
The scattered light is collected with a 50 \textmu m core optical
fiber connected to a 50 cm grating spectrometer equipped with a
nitrogen cooled Charge Coupled Device (CCD) camera. The sample is
mounted on piezo stages allowing to move the sample under the
laser spot with a sub-\textmu m resolution. Spatial maps of the
scattered intensity at a given energy (or energy interval) have
been performed by moving the sample under the laser spot and
collecting, at each point, the Raman scattering spectrum.
Magnetic field up to 14 T was produced by a superconducting coil.

\begin{figure}
\begin{center}
\includegraphics[width=0.9\linewidth,angle=0,clip]{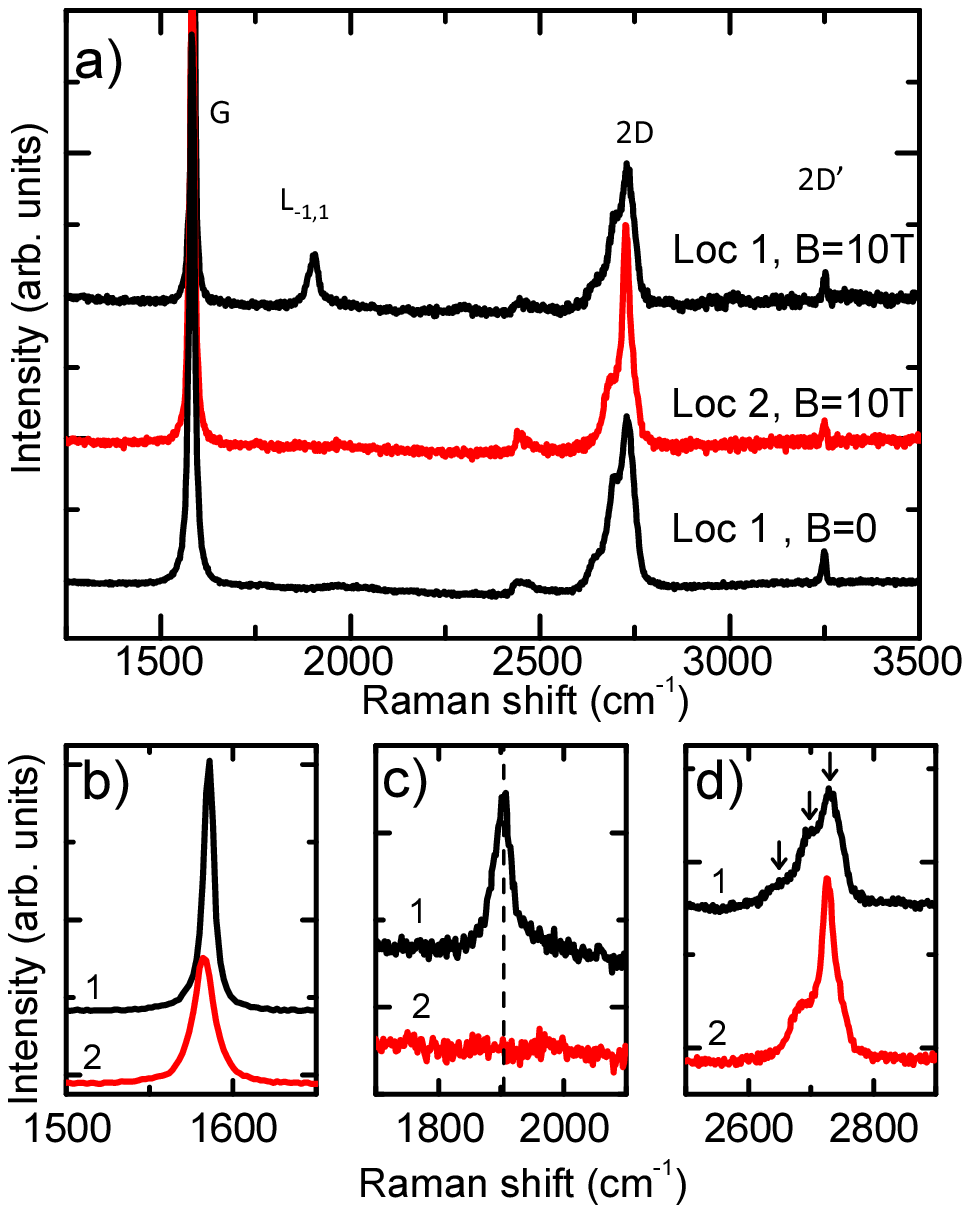}
\caption{(Color online) a) Typical Raman scattering spectra at
two different locations on the surface of bulk graphite measured
at $T=4.2$~K and both at $B=0$ and at $B=10$~T using $\lambda$=514.5~nm excitation.
 b,c,d) details of the same spectra in the range of energy of the G
band, of the electronic excitation L$_{-1,1}$ at $B=10$~T, and of the 2D band,
respectively. Spectra have been shifted vertically for clarity.}
\label{Fig11}
\end{center}
\end{figure}

Graphene and bulk graphite present distinct electronic excitation spectra. When a
magnetic field is applied, Landau levels in graphene are formed
with an evolution with magnetic field defined as
$E_n=$sign($n$)$v_F\sqrt{2 e \hbar B |n|}$, where $v_F$ is the
Fermi velocity and n the Landau level index. Electronic excitations between Landau
levels n and m are labelled $L_{m,n}=E_n-E_m$, and their energy grows like $\sqrt{B}$.
The electronic contribution to the Raman scattering response of bulk graphite is dominated by carriers at
the K point of the Brillouin zone~\cite{Kossacki2011}. Their
excitation spectrum is similar to the one of graphene bilayer, with a linear evolution
of the Landau level energies with increasing magnetic fields, but
with an effective inter-layer coupling constant $\gamma_1^*$ twice
enhanced~\cite{Koshino2008,Orlita2009}.

The methodology proposed here is based on the spatial mapping of the unpolarized Raman scattering response
of the surface of bulk graphite with an applied magnetic field, in this case, $B=10$~T.
Both the phonon and the purely electronic
contributions to the spectra are compared for different locations.
We have determined three distinct criteria that can be used to identify the
graphene-like locations.

Fig.~\ref{Fig11}a) shows two Raman scattering spectra measured at $B=10$~T at two different
locations on the surface of bulk graphite, with an excitation laser
at $\lambda=514.5$~nm.

The spectrum measured at location 2, is typical of bulk graphite. The G band appears at
$1582$~cm$^{-1}$  with a full width at half maximum (FWHM) of
$14$~cm$^{-1}$ and the second order peaks such as the 2D ($\sim2700$~cm$^{-1}$) and the
2D' ($\sim3250$~cm$^{-1}$) bands.

The measured spectrum is rather different when placing
the laser spot at the location 1, and it is typical of a graphene
flake on bulk graphite. The G band measured at $B=10$~T  now appears at $1586$~cm$^{-1}$
with a FWHM of $7.5$~cm$^{-1}$. This difference can be better
observed in Fig.~\ref{Fig11}b). The blue shift of the G band
together with the strong reduction of its FWHM and the increase of its amplitude,
not observed at $B=0$~T, are due to the magneto-phonon resonance~\cite{Ando2007,Goerbig2007}, the
 resonant coupling of optical phonons with magneto-excitons, which is much more
pronounced in graphene~\cite{Faugeras2009} than in bulk
graphite~\cite{Kossacki2011,Kim2012} at this particular value of
the magnetic field. Then, an additional feature at $\sim
1904$~cm$^{-1}$ is observed in the spectrum measured at location
1. This new excitation, only observed with an applied magnetic field and when the laser spot is
located on the flake, is the electronic inter Landau level
excitation~\cite{Faugeras2011} $L_{-1,1}$, between Landau levels
of index $n=-1$ and $m=1$. This particular excitation is expected
to be the most pronounced feature in the electronic contribution
to the Raman scattering response of graphene in a magnetic
field~\cite{Kashuba2009}, and is not observed in bulk graphite. This feature is more clearly shown in Fig.~\ref{Fig11}c).
Finally, as can be seen in Fig.~\ref{Fig11}d), the 2D band also acquires a fine structure.
When measured with the $514.5$~nm line of an Ar$^+$ laser, it is
composed of three components centered at 2732, 2695 and
2651~cm$^{-1}$ (indicated by arrows in Fig.~\ref{Fig11}d). The 2D
band of bulk graphite, measured in the same conditions, presents
two components centered at 2726 and 2684~cm$^{-1}$. As the 2D band
feature is only weakly affected by a magnetic
field~\cite{Faugeras2010a}, this criterium is also valid at $B=0$ (see spectrum in Fig.~\ref{Fig11}a).
One can note that the 2D band at location 1 is not simply the superposition
of a graphite-like and graphene-like 2D bands, but shows that the electronic band structure, mostly responsible for the shape of the 2D band~\cite{Ferrari2006}, is modified. The observed 2D band shape appears
rather different than the one routinely measured on twisted bilayer graphene
which appears as a single component lorentzian shaped feature~\cite{Ni2008}.
A characteristic Raman feature of twisted bilayers is the R band, with an energy which depends on the
twist angle between the two graphene layers~\cite{Carozo2011,He2013}.
In the present experiment, no R band is observed at any locations
on the investigated surfaces.

\begin{figure}
\begin{center}
\includegraphics[width=0.7\linewidth,angle=0,clip]{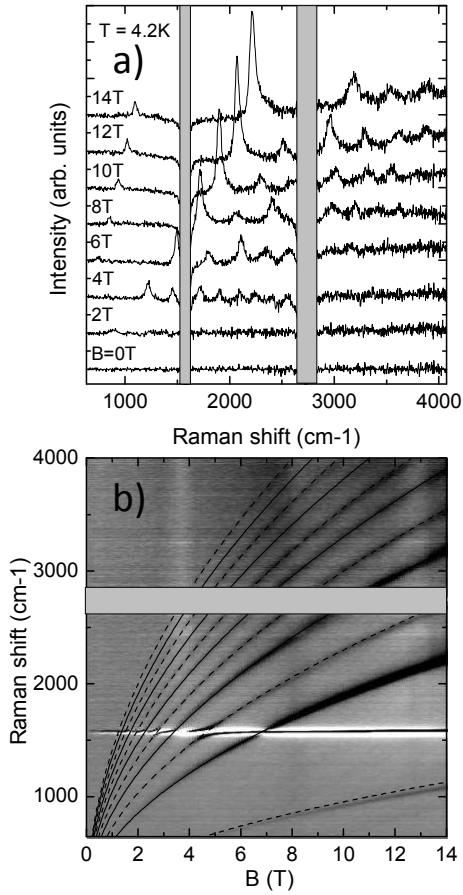}
\caption{a) Typical Raman scattering spectra from which the $B=0$~T
spectrum has been subtracted, measured at
different values of the magnetic field on a graphene on graphite
location measured with $\lambda$=720~nm excitation. b) Gray scale
map of the magnetic evolution of the Raman scattering response.
The dashed and solid lines represent $\Delta|n|=\pm 1$ and
$\Delta|n|=0$ inter Landau level excitations, respectively,
calculated using a Fermi velocity
$v_F=(1.03\pm0.01)\times10^{6}$~m.s$^{-1}$.} \label{Fig21}
\end{center}
\end{figure}

When placing the laser spot on a graphene-like location and increasing
the magnetic field, many different dispersing features can be observed
in the unpolarized Raman scattering response (see Fig.~\ref{Fig21}a)).
They can be attributed to inter Landau level electronic excitations, and
their observation can be used to determine the Fermi velocity following
$v_F=(E_n-E_m)/(\sqrt{2 e \hbar B (|n|+|m|))}$. Fig.~\ref{Fig21}b) shows
a gray scale map of the evolution of these excitations as a function of
the magnetic field, together with dashed and solid lines calculated with
$v_F=(1.03\pm0.01)\times10^{6}$~m.s$^{-1}$.

\begin{figure}
\begin{center}
\includegraphics[width=0.7\linewidth,angle=0,clip]{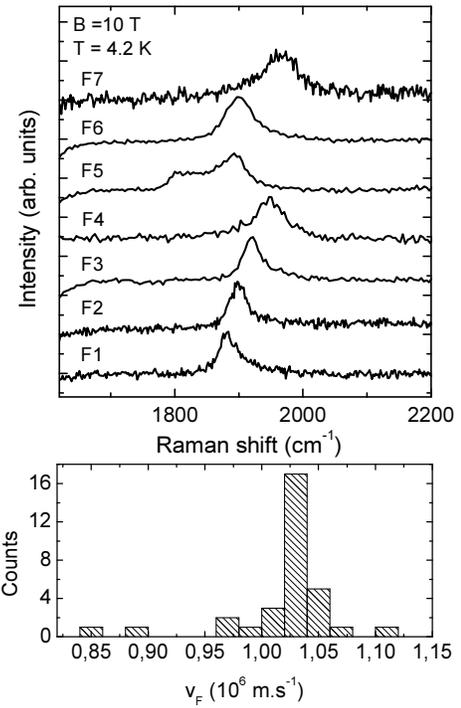}
\caption{a) L$_{-1,1}$ Raman scattering feature measured at $B =
10$~T on 7 different decoupled graphene flake locations on the
surface of bulk graphite (F1 to F7). b) Histogram of the
distribution of observed Fermi velocities out of 25 different flakes.}\label{Fig31}
\end{center}
\end{figure}

The average Fermi velocity depends on different parameters such as
the dielectric constant of the underlying
substrate~\cite{Hwang2012}, on possible strains affecting the
lattice parameter or, in the case of twisted graphene bilayers, on
the twist angle between the two layers~\cite{Lopes07,Trambly2010}.
Fig.~\ref{Fig31}a) shows the $L_{-1,1}$ feature measured at
$B=10$~T for 7 different graphene flakes on bulk graphite. The
energies at which this feature is observed at $B=10$~T range from
$1800$ to $2000$~cm$^{-1}$ with pronounced variations of its FWHM
and of its line shape. For instance, $F5$ shows a double peak
feature, similar to the one observed by Kuhne \textit{et
al.}\cite{Kuhne2012}, probably arising from two distinct graphene
flakes. We have measured more than 25 different flakes and, in
Fig.~\ref{Fig31}b), we present a histogram showing the different
values of Fermi velocities. The distribution is centered around
$v_F=1.025 \times 10^{6}$~m.s$^{-1}$ but observed $v_F$ can range
between $0.85\times10^{6}$ and $1.12\times10^{6}$~m.s$^{-1}$. As
all these flakes are on a similar substrate, dielectric screening
or strain effect should not differ from one flake to another. The
most likely reason for this observation is in the twist angle
between the decoupled flake and the substrate.

\begin{figure}
\begin{center}
\includegraphics[width=0.9\linewidth,angle=0,clip]{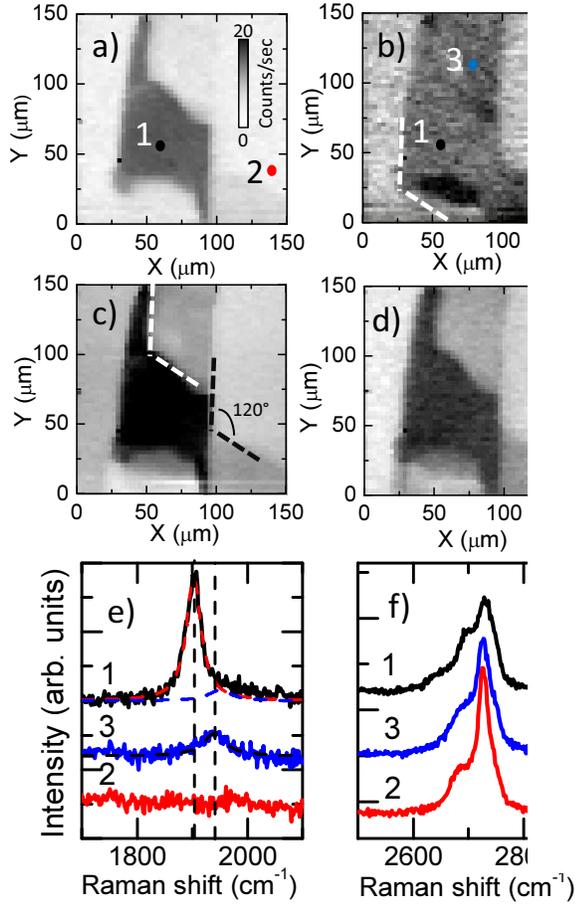}
\caption{(Color online) Gray scale spatial map with steps of
$3$~\textmu m measured at $B=10$~T of a) the scattered amplitude at $1904$~cm$^{-1}$
and indicating the two locations 1 and 2, b) the scattered
amplitude at $1940$~cm$^{-1}$ and indicating the two location 1
and 3, c) the scattered amplitudes at $1586$~cm$^{-1}$, and d) the
ratio of scattered amplitude
I($2695$~cm$^{-1}$)/I($2732$~cm$^{-1}$). Some angles of
$120^{\circ}$ are indicated by the dashed lines. e) Spectra in the range of the $L_{-1,1}$ feature at $B=10$~T for locations 1-3 indicated in panel a) and b). Dashed lines are lorentzian fits. f) Corresponding 2D band features.}
\label{Fig41}
\end{center}
\end{figure}

In Fig.~\ref{Fig41}a), we show the spatial variations of the scattered amplitude at $\sim 1904$~cm$^{-1}$,
for the same spatial region from which spectra presented in Fig.~\ref{Fig11} were acquired. This energy corresponds
to the expected energy of the$L_{-1,1}$ excitation at $B=10$~T. A large
region extending over more than $50$~\textmu m, with a well
defined shape and angles at $120^{\circ}$ typical of sp$^2$
carbon, is revealed. Fig.~\ref{Fig41}c) is a gray scale map of
the spatial variations of the scattered amplitude at
$1586$~cm$^{-1}$, at the energy of the coupled
phonon-magneto-exciton mode in graphene at $B=10$~T. The observed
shape in this figure is exactly the same as the one defined in
Fig.~\ref{Fig41}a), showing a correlation between the observation of
the particular L$_{-1,1}$ electronic excitation and of the blue shifted G band at $B=10$~T.
 Fig.~\ref{Fig41}d) is an illustration of how such locations can be detected even at $B=0$. It shows a gray scale
map of the spatial variations of the ratio of the scattered
amplitudes at $2695$~cm$^{-1}$ and at $2732$~cm$^{-1}$, the energies of two characteristic
components of the 2D band observed in Fig.~\ref{Fig11}d). The spatial region that
appears corresponds to the one defined in Fig.~\ref{Fig41}a) and
c) and defines again the decoupled graphene flake. All these three criteria can be used to visualize the shape of the
location giving rise to the graphene-like excitation spectrum.

A detailed investigation of the spectra measured in the scanning
area shows that, in the present case, we are not studying a single
graphene flake but at least two superimposed decoupled flakes.
This can be seen in Fig.~\ref{Fig41}e) by comparing the spectra
measured at locations 1 and 3, both showing a $L_{-1,1}$
electronic excitation but at slightly different energies. At
location 1, the electronic Raman scattering signal appears at
$\sim 1904$~cm$^{-1}$ which corresponds to $v_F=(1.029\pm0.003)
\times 10^6$~m.s$^{-1}$ with a FWHM of $\sim 31$~cm$^{-1}$ at
$B=10$~T. At location 3, this feature appears at $\sim
1940$~cm$^{-1}$ ($v_F=(1.05\pm0.005) \times 10^6$~m.s$^{-1}$)
with a FWHM of $54$~cm$^{-1}$. The integrated intensity of the
$L_{-1,1}$ excitation at location 3 is three times smaller than
the one measured at location 1. The reason for this difference in
the measured intensities at these two different locations remains
unknown. Both features have an energy increasing like $\sqrt{B}$, as
is expected for a graphene-like excitation spectrum. The G band feature at location 3
appears at $\sim 1584$~cm$^{-1}$ with a FWHM of $10$~cm$^{-1}$, in
between the positions of G band observed at location 1 and 2, indicating a less
efficient electron-phonon interaction than at location 1. Even
though we have no clear interpretation for this observation, it is
in line with the FWHM for the L$_{-1,1}$ feature nearly twice
bigger at location 3 than at location 1. To show the spatial
extent of the regions giving rise to this additional signal, we
have built a spatial map of the scattered intensity at
$1940$~cm$^{-1}$ which is presented in Figure~\ref{Fig41}b). As
can be seen, this region is much larger than the one defined in
Fig.~\ref{Fig41}a), it has also a well defined shape, and
describes a distinct flake, characterized by a Fermi velocity $2\%$ higher in this case. Changes in the Fermi velocity could be
accounted for by considering that the first layer presents a
different rotation angle with the bulk graphite substrate than the
second layer with the first layer.

\begin{figure}
\begin{center}
\includegraphics[width=0.9\linewidth,angle=0,clip]{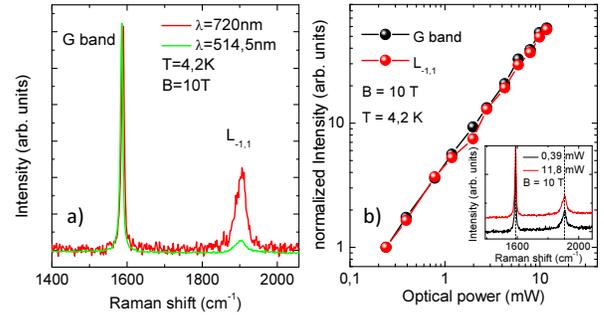}
\caption{(Color online) a) Typical Raman scattering spectrum of
the graphene flake normalized to the G band intensity, measured at
$B=10$~T and at $T=4.2$~K, using $\lambda=$514.5~nm (green
spectrum) and $\lambda=$720~nm excitation (red spectrum). b)
Excitation power dependence, measured with $\lambda=720 nm$, of the integrated intensity of the G
band feature (black dots) and of the $L_{-1,1}$ feature (red
dots), normalized to their intensity measured at the lowest power,
and measured at $T=4.2$~K. Inset: Spectra measured with 0.39 mW (black) and with 11.8 mW (red) at $B=10$~T, normalized to excitation power and to the acquisition time.} \label{Fig51}
\end{center}
\end{figure}

To further investigate this system, we have performed similar
experiments with a $\lambda=720$~nm excitation provided by a
Ti:Sapphire laser. We were able to locate the same graphene flake
on the bulk graphite substrate as the one shown in Fig.~\ref{Fig41} and to compare the Raman scattering
responses for these two excitation energies. As we do not have access experimentally to the absolute scattered intensity, we can normalize the measured spectra to the G band intensity and compare the relative intensities of the L$_{-1,1}$ features measured with two different excitation wavelengths. Figure~\ref{Fig51}a) shows
two Raman scattering spectra of the G band and
of the $L_{-1,1}$ features, normalized to the G band intensity, measured on the same graphene flake on
bulk graphite at $T=4.2$~K and at $B=10$~T. It appears clearly that the
$L_{-1,1}$ feature is more pronounced when using $\lambda=720$~nm
than $\lambda=514.5$~nm excitation. We find that the integrated
intensity ratio is $\sim 5.6$. One can also note that
 we do not observe any change in the energy of this excitation
 with the excitation laser energy.

The integrated intensities of both the phonon and of the electronic
excitation features are expected to vary with the excitation laser
energy. Concerning the electronic $L_{-1,1}$ feature, a recent
theoretical work~\cite{Kashuba2009} has proposed that the
scattered intensity should scale as $1/\Omega^2$ where $\Omega$ is
the laser excitation energy and these excitations are thus
expected to be more intense when using high wavelength excitation.
$L_{-1,1}$ is expected to be 1.96 times more intense when measured
at $\lambda=720$~nm than when measured at $\lambda=514.5$~nm. On the
other hand, because of the different type of Raman processes, the
scattered intensity by optical phonons has been shown~\cite{Pimenta2007,Basko2009} to scale
as $\Omega^2.f(\Omega/t_0)$ where
$\Omega$ is the excitation photon frequency, $t_0$ the nearest
neighbors hopping integral ($\sim 3~$eV) and $f(x)$ is a growing,
positive and dimensionless function evaluated in
Ref.~\cite{Basko2009}. The G band feature is hence more pronounced
using low wavelength excitations. Evaluating $f(x)$ to be $\sim
1.1$ for $\lambda=720$~nm and to be $\sim 1.4$ for
$\lambda=514.5$~nm, one expects the G band to be 2.44 times more
intense when measured with $\lambda=514.5$~nm than with
$\lambda=720$~nm. From the two theoretical works by O. Kashuba
\textit{et al.}~\cite{Kashuba2009} and by D.M.
Basko~\cite{Basko2009}, after normalizing the obtained spectra to
the G band intensity, one expects the integrated intensity of the
$L_{-1,1}$ feature measured at $\lambda=720$~nm to be 4.8 times
larger than the one measured at $\lambda=514.5$~nm, in good agreement
with the experimentally observed evolution.

Figure~\ref{Fig51}b) shows the evolution of the integrated intensity as a function of the
excitation power of both the G band and the $L_{-1,1}$ features, normalized by their value measured at the
lowest power level. As can be seen in this figure, up to $10$~mW
of excitation power focused on $\sim 1$~\textmu m excitation spot,
the intensity of these two features is growing linearly with the
excitation power, with no observed changes in their FWHM nor in
their energies (see spectra in the inset of Fig.~\ref{Fig51}b). Because a strong excitation power dependence of the phonon
frequency has been observed for suspended graphene and not for
supported graphene~\cite{Cai2010}, this result shows that the graphene flake is
in very good thermal contact with the underlying graphite
substrate.

To conclude, we have presented a set of magneto-Raman scattering
experiments performed on graphene flakes on the surface of bulk graphite.
Based on their specific Raman scattering spectra, we have reported on three different ways to identify
these locations. Our results confirm
the exceptionally high quality of the electronic states with the
observation of features associated with electronic excitations of
FWHM~$\sim 30$~cm$^{-1}$ at $B=10$~T. We show a small spreading in the
measured Fermi velocities of different graphene flakes. These experiments allow for the
experimental demonstration of the excitation wavelength dependence
of both phonon and electronic excitations intensity in the Raman
scattering response of graphene, in agreement with recent
theories.

\acknowledgments We acknowledge fruitful discussions with D.M.
Basko. The work of J.B. was supported by the Foundation for Polish Science International PhD Projects Programme co-financed by the EU European Regional Development Fund. Part of this work has been supported by the graphene
flagship project, by the European Research Council
(ERC-2012-AdG-320590-MOMB) and by the Polish Center for Scientific Research (NCN) project DEC-2013/10/M/ST3/00791.


\begin{thebibliography}{35}%
\makeatletter
\providecommand \@ifxundefined [1]{%
 \@ifx{#1\undefined}
}%
\providecommand \@ifnum [1]{%
 \ifnum #1\expandafter \@firstoftwo
 \else \expandafter \@secondoftwo
 \fi
}%
\providecommand \@ifx [1]{%
 \ifx #1\expandafter \@firstoftwo
 \else \expandafter \@secondoftwo
 \fi
}%
\providecommand \natexlab [1]{#1}%
\providecommand \enquote  [1]{``#1''}%
\providecommand \bibnamefont  [1]{#1}%
\providecommand \bibfnamefont [1]{#1}%
\providecommand \citenamefont [1]{#1}%
\providecommand \href@noop [0]{\@secondoftwo}%
\providecommand \href [0]{\begingroup \@sanitize@url \@href}%
\providecommand \@href[1]{\@@startlink{#1}\@@href}%
\providecommand \@@href[1]{\endgroup#1\@@endlink}%
\providecommand \@sanitize@url [0]{\catcode `\\12\catcode
`\$12\catcode
  `\&12\catcode `\#12\catcode `\^12\catcode `\_12\catcode `\%12\relax}%
\providecommand \@@startlink[1]{}%
\providecommand \@@endlink[0]{}%
\providecommand \url  [0]{\begingroup\@sanitize@url \@url }%
\providecommand \@url [1]{\endgroup\@href {#1}{\urlprefix }}%
\providecommand \urlprefix  [0]{URL }%
\providecommand \Eprint [0]{\href }%
\providecommand \doibase [0]{http://dx.doi.org/}%
\providecommand \selectlanguage [0]{\@gobble}%
\providecommand \bibinfo  [0]{\@secondoftwo}%
\providecommand \bibfield  [0]{\@secondoftwo}%
\providecommand \translation [1]{[#1]}%
\providecommand \BibitemOpen [0]{}%
\providecommand \bibitemStop [0]{}%
\providecommand \bibitemNoStop [0]{.\EOS\space}%
\providecommand \EOS [0]{\spacefactor3000\relax}%
\providecommand \BibitemShut  [1]{\csname bibitem#1\endcsname}%
\let\auto@bib@innerbib\@empty
\bibitem [{\citenamefont {Novoselov}\ \emph {et~al.}(2005)\citenamefont
  {Novoselov}, \citenamefont {Geim}, \citenamefont {Morozov}, \citenamefont
  {Jiang}, \citenamefont {Katsnelson}, \citenamefont {Grigorieva},
  \citenamefont {Dubonos},\ and\ \citenamefont {Firsov}}]{Novoselov2005}%
  \BibitemOpen
  \bibfield  {author} {\bibinfo {author} {\bibfnamefont {K.}~\bibnamefont
  {Novoselov}}, \bibinfo {author} {\bibfnamefont {A.}~\bibnamefont {Geim}},
  \bibinfo {author} {\bibfnamefont {S.}~\bibnamefont {Morozov}}, \bibinfo
  {author} {\bibfnamefont {D.}~\bibnamefont {Jiang}}, \bibinfo {author}
  {\bibfnamefont {M.}~\bibnamefont {Katsnelson}}, \bibinfo {author}
  {\bibfnamefont {I.}~\bibnamefont {Grigorieva}}, \bibinfo {author}
  {\bibfnamefont {S.}~\bibnamefont {Dubonos}}, \ and\ \bibinfo {author}
  {\bibfnamefont {A.}~\bibnamefont {Firsov}},\ }\href@noop {} {\bibfield
  {journal} {\bibinfo  {journal} {Nature}\ }\textbf {\bibinfo {volume} {438}},\
  \bibinfo {pages} {197} (\bibinfo {year} {2005})}\BibitemShut {NoStop}%
\bibitem [{\citenamefont {Zhang}\ \emph {et~al.}(2005)\citenamefont {Zhang},
  \citenamefont {Tan}, \citenamefont {Stormer},\ and\ \citenamefont
  {Kim}}]{Zhang2005}%
  \BibitemOpen
  \bibfield  {author} {\bibinfo {author} {\bibfnamefont {Y.}~\bibnamefont
  {Zhang}}, \bibinfo {author} {\bibfnamefont {Y.}~\bibnamefont {Tan}}, \bibinfo
  {author} {\bibfnamefont {H.}~\bibnamefont {Stormer}}, \ and\ \bibinfo
  {author} {\bibfnamefont {P.}~\bibnamefont {Kim}},\ }\href@noop {} {\bibfield
  {journal} {\bibinfo  {journal} {Nature}\ }\textbf {\bibinfo {volume} {438}},\
  \bibinfo {pages} {201} (\bibinfo {year} {2005})}\BibitemShut {NoStop}%
\bibitem [{\citenamefont {Martin}\ \emph {et~al.}(2008)\citenamefont {Martin},
  \citenamefont {Akerman}, \citenamefont {Ulbricht}, \citenamefont {Lohmann},
  \citenamefont {Smet}, \citenamefont {Von~Klitzing},\ and\ \citenamefont
  {Yacoby}}]{Martin2008}%
  \BibitemOpen
  \bibfield  {author} {\bibinfo {author} {\bibfnamefont {J.}~\bibnamefont
  {Martin}}, \bibinfo {author} {\bibfnamefont {N.}~\bibnamefont {Akerman}},
  \bibinfo {author} {\bibfnamefont {G.}~\bibnamefont {Ulbricht}}, \bibinfo
  {author} {\bibfnamefont {T.}~\bibnamefont {Lohmann}}, \bibinfo {author}
  {\bibfnamefont {J.~H.}\ \bibnamefont {Smet}}, \bibinfo {author}
  {\bibfnamefont {K.}~\bibnamefont {Von~Klitzing}}, \ and\ \bibinfo {author}
  {\bibfnamefont {A.}~\bibnamefont {Yacoby}},\ }\href {\doibase
  10.1038/nphys781} {\bibfield  {journal} {\bibinfo  {journal} {Nat. Phys.}\
  }\textbf {\bibinfo {volume} {4}},\ \bibinfo {pages} {144} (\bibinfo {year}
  {2008})}\BibitemShut {NoStop}%
\bibitem [{\citenamefont {Bolotin}\ \emph {et~al.}(2008)\citenamefont
  {Bolotin}, \citenamefont {Sikes}, \citenamefont {Jiang}, \citenamefont
  {Klima}, \citenamefont {Fudenberg}, \citenamefont {Hone}, \citenamefont
  {Kim},\ and\ \citenamefont {Stormer}}]{Bolotin2008}%
  \BibitemOpen
  \bibfield  {author} {\bibinfo {author} {\bibfnamefont {K.~I.}\ \bibnamefont
  {Bolotin}}, \bibinfo {author} {\bibfnamefont {K.~J.}\ \bibnamefont {Sikes}},
  \bibinfo {author} {\bibfnamefont {Z.}~\bibnamefont {Jiang}}, \bibinfo
  {author} {\bibfnamefont {M.}~\bibnamefont {Klima}}, \bibinfo {author}
  {\bibfnamefont {G.}~\bibnamefont {Fudenberg}}, \bibinfo {author}
  {\bibfnamefont {J.}~\bibnamefont {Hone}}, \bibinfo {author} {\bibfnamefont
  {P.}~\bibnamefont {Kim}}, \ and\ \bibinfo {author} {\bibfnamefont {H.~L.}\
  \bibnamefont {Stormer}},\ }\href@noop {} {\bibfield  {journal} {\bibinfo
  {journal} {Solid State Comm.}\ }\textbf {\bibinfo {volume} {146}},\ \bibinfo
  {pages} {351} (\bibinfo {year} {2008})}\BibitemShut {NoStop}%
\bibitem [{\citenamefont {Berciaud}\ \emph {et~al.}(2009)\citenamefont
  {Berciaud}, \citenamefont {Ryu}, \citenamefont {Brus},\ and\ \citenamefont
  {Heinz}}]{Berciaud2009}%
  \BibitemOpen
  \bibfield  {author} {\bibinfo {author} {\bibfnamefont {S.}~\bibnamefont
  {Berciaud}}, \bibinfo {author} {\bibfnamefont {S.}~\bibnamefont {Ryu}},
  \bibinfo {author} {\bibfnamefont {L.~E.}\ \bibnamefont {Brus}}, \ and\
  \bibinfo {author} {\bibfnamefont {T.~F.}\ \bibnamefont {Heinz}},\ }\href@noop
  {} {\bibfield  {journal} {\bibinfo  {journal} {NanoLett.}\ }\textbf {\bibinfo
  {volume} {9}},\ \bibinfo {pages} {346} (\bibinfo {year} {2009})}\BibitemShut
  {NoStop}%
\bibitem [{\citenamefont {Ki}\ \emph {et~al.}(2013)\citenamefont {Ki},
  \citenamefont {Fal'ko},\ and\ \citenamefont {Morpurgo}}]{Ki2013}%
  \BibitemOpen
  \bibfield  {author} {\bibinfo {author} {\bibfnamefont {D.-K.}\ \bibnamefont
  {Ki}}, \bibinfo {author} {\bibfnamefont {V.~I.}\ \bibnamefont {Fal'ko}}, \
  and\ \bibinfo {author} {\bibfnamefont {A.~F.}\ \bibnamefont {Morpurgo}},\
  }\href@noop {} {} (\bibinfo {year} {2013}),\ \Eprint
  {http://arxiv.org/abs/condmat/1305.4761} {arXiv:condmat/1305.4761}
  \BibitemShut {NoStop}%
\bibitem [{\citenamefont {Dean}\ \emph {et~al.}(2010)\citenamefont {Dean},
  \citenamefont {Young}, \citenamefont {Meric}, \citenamefont {Lee},
  \citenamefont {Wang}, \citenamefont {Sorgenfrei}, \citenamefont {Watanabe},
  \citenamefont {Taniguchi}, \citenamefont {Kim}, \citenamefont {Shepard},\
  and\ \citenamefont {Hone}}]{Dean2010}%
  \BibitemOpen
  \bibfield  {author} {\bibinfo {author} {\bibfnamefont {C.~R.}\ \bibnamefont
  {Dean}}, \bibinfo {author} {\bibfnamefont {A.~F.}\ \bibnamefont {Young}},
  \bibinfo {author} {\bibfnamefont {I.}~\bibnamefont {Meric}}, \bibinfo
  {author} {\bibfnamefont {C.}~\bibnamefont {Lee}}, \bibinfo {author}
  {\bibfnamefont {L.}~\bibnamefont {Wang}}, \bibinfo {author} {\bibfnamefont
  {S.}~\bibnamefont {Sorgenfrei}}, \bibinfo {author} {\bibfnamefont
  {K.}~\bibnamefont {Watanabe}}, \bibinfo {author} {\bibfnamefont
  {T.}~\bibnamefont {Taniguchi}}, \bibinfo {author} {\bibfnamefont
  {P.}~\bibnamefont {Kim}}, \bibinfo {author} {\bibfnamefont {K.~L.}\
  \bibnamefont {Shepard}}, \ and\ \bibinfo {author} {\bibfnamefont
  {J.}~\bibnamefont {Hone}},\ }\href@noop {} {\bibfield  {journal} {\bibinfo
  {journal} {Nature Nanotech.}\ }\textbf {\bibinfo {volume} {5}},\ \bibinfo
  {pages} {722} (\bibinfo {year} {2010})}\BibitemShut {NoStop}%
\bibitem [{\citenamefont {Bolotin}\ \emph {et~al.}(2009)\citenamefont
  {Bolotin}, \citenamefont {Ghahari}, \citenamefont {Shulman}, \citenamefont
  {Stormer},\ and\ \citenamefont {Kim}}]{Bolotin2009}%
  \BibitemOpen
  \bibfield  {author} {\bibinfo {author} {\bibfnamefont {K.~I.}\ \bibnamefont
  {Bolotin}}, \bibinfo {author} {\bibfnamefont {F.}~\bibnamefont {Ghahari}},
  \bibinfo {author} {\bibfnamefont {M.~D.}\ \bibnamefont {Shulman}}, \bibinfo
  {author} {\bibfnamefont {H.~L.}\ \bibnamefont {Stormer}}, \ and\ \bibinfo
  {author} {\bibfnamefont {P.}~\bibnamefont {Kim}},\ }\href@noop {} {\bibfield
  {journal} {\bibinfo  {journal} {Nature}\ }\textbf {\bibinfo {volume} {462}},\
  \bibinfo {pages} {196} (\bibinfo {year} {2009})}\BibitemShut {NoStop}%
\bibitem [{\citenamefont {Du}\ \emph {et~al.}(2009)\citenamefont {Du},
  \citenamefont {Skachko}, \citenamefont {Duerr}, \citenamefont {Luican},\ and\
  \citenamefont {Andrei}}]{Du2009}%
  \BibitemOpen
  \bibfield  {author} {\bibinfo {author} {\bibfnamefont {X.}~\bibnamefont
  {Du}}, \bibinfo {author} {\bibfnamefont {I.}~\bibnamefont {Skachko}},
  \bibinfo {author} {\bibfnamefont {F.}~\bibnamefont {Duerr}}, \bibinfo
  {author} {\bibfnamefont {A.}~\bibnamefont {Luican}}, \ and\ \bibinfo {author}
  {\bibfnamefont {E.~Y.}\ \bibnamefont {Andrei}},\ }\href@noop {} {\bibfield
  {journal} {\bibinfo  {journal} {Nature}\ }\textbf {\bibinfo {volume} {462}},\
  \bibinfo {pages} {192} (\bibinfo {year} {2009})}\BibitemShut {NoStop}%
\bibitem [{\citenamefont {Dean}\ \emph {et~al.}(2011)\citenamefont {Dean},
  \citenamefont {Young}, \citenamefont {Cadden-Zimansky}, \citenamefont {Wang},
  \citenamefont {Ren}, \citenamefont {Watanabe}, \citenamefont {Taniguchi},
  \citenamefont {Kim}, \citenamefont {Hone},\ and\ \citenamefont
  {Shepard}}]{Dean2011}%
  \BibitemOpen
  \bibfield  {author} {\bibinfo {author} {\bibfnamefont {C.~R.}\ \bibnamefont
  {Dean}}, \bibinfo {author} {\bibfnamefont {A.~F.}\ \bibnamefont {Young}},
  \bibinfo {author} {\bibfnamefont {P.}~\bibnamefont {Cadden-Zimansky}},
  \bibinfo {author} {\bibfnamefont {L.}~\bibnamefont {Wang}}, \bibinfo {author}
  {\bibfnamefont {H.}~\bibnamefont {Ren}}, \bibinfo {author} {\bibfnamefont
  {K.}~\bibnamefont {Watanabe}}, \bibinfo {author} {\bibfnamefont
  {T.}~\bibnamefont {Taniguchi}}, \bibinfo {author} {\bibfnamefont
  {P.}~\bibnamefont {Kim}}, \bibinfo {author} {\bibfnamefont {J.}~\bibnamefont
  {Hone}}, \ and\ \bibinfo {author} {\bibfnamefont {K.~L.}\ \bibnamefont
  {Shepard}},\ }\href@noop {} {\bibfield  {journal} {\bibinfo  {journal}
  {Nature Phys.}\ }\textbf {\bibinfo {volume} {7}},\ \bibinfo {pages} {693}
  (\bibinfo {year} {2011})}\BibitemShut {NoStop}%
\bibitem [{\citenamefont {Li}\ \emph {et~al.}(2009)\citenamefont {Li},
  \citenamefont {Luican},\ and\ \citenamefont {Andrei}}]{Li2009}%
  \BibitemOpen
  \bibfield  {author} {\bibinfo {author} {\bibfnamefont {G.}~\bibnamefont
  {Li}}, \bibinfo {author} {\bibfnamefont {A.}~\bibnamefont {Luican}}, \ and\
  \bibinfo {author} {\bibfnamefont {E.~Y.}\ \bibnamefont {Andrei}},\
  }\href@noop {} {\bibfield  {journal} {\bibinfo  {journal} {Phys. Rev. Lett.}\
  }\textbf {\bibinfo {volume} {102}},\ \bibinfo {pages} {176804} (\bibinfo
  {year} {2009})}\BibitemShut {NoStop}%
\bibitem [{\citenamefont {Neugebauer}\ \emph {et~al.}(2009)\citenamefont
  {Neugebauer}, \citenamefont {Orlita}, \citenamefont {Faugeras}, \citenamefont
  {Barra},\ and\ \citenamefont {Potemski}}]{Neugebauer2009}%
  \BibitemOpen
  \bibfield  {author} {\bibinfo {author} {\bibfnamefont {P.}~\bibnamefont
  {Neugebauer}}, \bibinfo {author} {\bibfnamefont {M.}~\bibnamefont {Orlita}},
  \bibinfo {author} {\bibfnamefont {C.}~\bibnamefont {Faugeras}}, \bibinfo
  {author} {\bibfnamefont {A.-L.}\ \bibnamefont {Barra}}, \ and\ \bibinfo
  {author} {\bibfnamefont {M.}~\bibnamefont {Potemski}},\ }\href@noop {}
  {\bibfield  {journal} {\bibinfo  {journal} {Phys. Rev. Lett.}\ }\textbf
  {\bibinfo {volume} {103}},\ \bibinfo {pages} {136403} (\bibinfo {year}
  {2009})}\BibitemShut {NoStop}%
\bibitem [{\citenamefont {Yan}\ \emph {et~al.}(2010)\citenamefont {Yan},
  \citenamefont {Goler}, \citenamefont {Rhone}, \citenamefont {Han},
  \citenamefont {He}, \citenamefont {Kim}, \citenamefont {Pellegrini},\ and\
  \citenamefont {Pinczuk}}]{Yan2010}%
  \BibitemOpen
  \bibfield  {author} {\bibinfo {author} {\bibfnamefont {J.}~\bibnamefont
  {Yan}}, \bibinfo {author} {\bibfnamefont {S.}~\bibnamefont {Goler}}, \bibinfo
  {author} {\bibfnamefont {T.~D.}\ \bibnamefont {Rhone}}, \bibinfo {author}
  {\bibfnamefont {M.}~\bibnamefont {Han}}, \bibinfo {author} {\bibfnamefont
  {R.}~\bibnamefont {He}}, \bibinfo {author} {\bibfnamefont {P.}~\bibnamefont
  {Kim}}, \bibinfo {author} {\bibfnamefont {V.}~\bibnamefont {Pellegrini}}, \
  and\ \bibinfo {author} {\bibfnamefont {A.}~\bibnamefont {Pinczuk}},\ }\href
  {\doibase 10.1103/PhysRevLett.105.227401} {\bibfield  {journal} {\bibinfo
  {journal} {Phys. Rev. Lett.}\ }\textbf {\bibinfo {volume} {105}},\ \bibinfo
  {pages} {227401} (\bibinfo {year} {2010})}\BibitemShut {NoStop}%
\bibitem [{\citenamefont {Faugeras}\ \emph {et~al.}(2011)\citenamefont
  {Faugeras}, \citenamefont {Amado}, \citenamefont {Kossacki}, \citenamefont
  {Orlita}, \citenamefont {K\"uhne}, \citenamefont {Nicolet}, \citenamefont
  {Latyshev},\ and\ \citenamefont {Potemski}}]{Faugeras2011}%
  \BibitemOpen
  \bibfield  {author} {\bibinfo {author} {\bibfnamefont {C.}~\bibnamefont
  {Faugeras}}, \bibinfo {author} {\bibfnamefont {M.}~\bibnamefont {Amado}},
  \bibinfo {author} {\bibfnamefont {P.}~\bibnamefont {Kossacki}}, \bibinfo
  {author} {\bibfnamefont {M.}~\bibnamefont {Orlita}}, \bibinfo {author}
  {\bibfnamefont {M.}~\bibnamefont {K\"uhne}}, \bibinfo {author} {\bibfnamefont
  {A.~A.~L.}\ \bibnamefont {Nicolet}}, \bibinfo {author} {\bibfnamefont
  {Y.~I.}\ \bibnamefont {Latyshev}}, \ and\ \bibinfo {author} {\bibfnamefont
  {M.}~\bibnamefont {Potemski}},\ }\href {\doibase
  10.1103/PhysRevLett.107.036807} {\bibfield  {journal} {\bibinfo  {journal}
  {Phys. Rev. Lett.}\ }\textbf {\bibinfo {volume} {107}},\ \bibinfo {pages}
  {036807} (\bibinfo {year} {2011})}\BibitemShut {NoStop}%
\bibitem [{\citenamefont {K\"uhne}\ \emph {et~al.}(2012)\citenamefont
  {K\"uhne}, \citenamefont {Faugeras}, \citenamefont {Kossacki}, \citenamefont
  {Nicolet}, \citenamefont {Orlita}, \citenamefont {Latyshev},\ and\
  \citenamefont {Potemski}}]{Kuhne2012}%
  \BibitemOpen
  \bibfield  {author} {\bibinfo {author} {\bibfnamefont {M.}~\bibnamefont
  {K\"uhne}}, \bibinfo {author} {\bibfnamefont {C.}~\bibnamefont {Faugeras}},
  \bibinfo {author} {\bibfnamefont {P.}~\bibnamefont {Kossacki}}, \bibinfo
  {author} {\bibfnamefont {A.~A.~L.}\ \bibnamefont {Nicolet}}, \bibinfo
  {author} {\bibfnamefont {M.}~\bibnamefont {Orlita}}, \bibinfo {author}
  {\bibfnamefont {Y.~I.}\ \bibnamefont {Latyshev}}, \ and\ \bibinfo {author}
  {\bibfnamefont {M.}~\bibnamefont {Potemski}},\ }\href {\doibase
  10.1103/PhysRevB.85.195406} {\bibfield  {journal} {\bibinfo  {journal} {Phys.
  Rev. B}\ }\textbf {\bibinfo {volume} {85}},\ \bibinfo {pages} {195406}
  (\bibinfo {year} {2012})}\BibitemShut {NoStop}%
\bibitem [{\citenamefont {Qiu}\ \emph {et~al.}(2013)\citenamefont {Qiu},
  \citenamefont {Shen}, \citenamefont {Cao}, \citenamefont {Cong},
  \citenamefont {Saito}, \citenamefont {Yu}, \citenamefont {Dresselhaus},\ and\
  \citenamefont {Yu}}]{Qiu2013}%
  \BibitemOpen
  \bibfield  {author} {\bibinfo {author} {\bibfnamefont {C.}~\bibnamefont
  {Qiu}}, \bibinfo {author} {\bibfnamefont {X.}~\bibnamefont {Shen}}, \bibinfo
  {author} {\bibfnamefont {B.}~\bibnamefont {Cao}}, \bibinfo {author}
  {\bibfnamefont {C.}~\bibnamefont {Cong}}, \bibinfo {author} {\bibfnamefont
  {R.}~\bibnamefont {Saito}}, \bibinfo {author} {\bibfnamefont
  {J.}~\bibnamefont {Yu}}, \bibinfo {author} {\bibfnamefont {M.~S.}\
  \bibnamefont {Dresselhaus}}, \ and\ \bibinfo {author} {\bibfnamefont
  {T.}~\bibnamefont {Yu}},\ }\href@noop {} {\bibfield  {journal} {\bibinfo
  {journal} {Phys. Rev. B}\ }\textbf {\bibinfo {volume} {88}},\ \bibinfo
  {pages} {165407} (\bibinfo {year} {2013})}\BibitemShut {NoStop}%
\bibitem [{\citenamefont {Kossacki}\ \emph {et~al.}(2011)\citenamefont
  {Kossacki}, \citenamefont {Faugeras}, \citenamefont {K\"uhne}, \citenamefont
  {Orlita}, \citenamefont {Nicolet}, \citenamefont {Schneider}, \citenamefont
  {Basko}, \citenamefont {Latyshev},\ and\ \citenamefont
  {Potemski}}]{Kossacki2011}%
  \BibitemOpen
  \bibfield  {author} {\bibinfo {author} {\bibfnamefont {P.}~\bibnamefont
  {Kossacki}}, \bibinfo {author} {\bibfnamefont {C.}~\bibnamefont {Faugeras}},
  \bibinfo {author} {\bibfnamefont {M.}~\bibnamefont {K\"uhne}}, \bibinfo
  {author} {\bibfnamefont {M.}~\bibnamefont {Orlita}}, \bibinfo {author}
  {\bibfnamefont {A.~A.~L.}\ \bibnamefont {Nicolet}}, \bibinfo {author}
  {\bibfnamefont {J.~M.}\ \bibnamefont {Schneider}}, \bibinfo {author}
  {\bibfnamefont {D.~M.}\ \bibnamefont {Basko}}, \bibinfo {author}
  {\bibfnamefont {Y.~I.}\ \bibnamefont {Latyshev}}, \ and\ \bibinfo {author}
  {\bibfnamefont {M.}~\bibnamefont {Potemski}},\ }\href {\doibase
  10.1103/PhysRevB.84.235138} {\bibfield  {journal} {\bibinfo  {journal} {Phys.
  Rev. B}\ }\textbf {\bibinfo {volume} {84}},\ \bibinfo {pages} {235138}
  (\bibinfo {year} {2011})}\BibitemShut {NoStop}%
\bibitem [{\citenamefont {Koshino}\ and\ \citenamefont
  {Ando}(2008)}]{Koshino2008}%
  \BibitemOpen
  \bibfield  {author} {\bibinfo {author} {\bibfnamefont {M.}~\bibnamefont
  {Koshino}}\ and\ \bibinfo {author} {\bibfnamefont {T.}~\bibnamefont {Ando}},\
  }\href@noop {} {\bibfield  {journal} {\bibinfo  {journal} {Phys. Rev. B}\
  }\textbf {\bibinfo {volume} {77}},\ \bibinfo {pages} {115313} (\bibinfo
  {year} {2008})}\BibitemShut {NoStop}%
\bibitem [{\citenamefont {Orlita}\ \emph {et~al.}(2009)\citenamefont {Orlita},
  \citenamefont {Faugeras}, \citenamefont {Schneider}, \citenamefont
  {Martinez}, \citenamefont {Maude},\ and\ \citenamefont
  {Potemski}}]{Orlita2009}%
  \BibitemOpen
  \bibfield  {author} {\bibinfo {author} {\bibfnamefont {M.}~\bibnamefont
  {Orlita}}, \bibinfo {author} {\bibfnamefont {C.}~\bibnamefont {Faugeras}},
  \bibinfo {author} {\bibfnamefont {J.}~\bibnamefont {Schneider}}, \bibinfo
  {author} {\bibfnamefont {G.}~\bibnamefont {Martinez}}, \bibinfo {author}
  {\bibfnamefont {D.~K.}\ \bibnamefont {Maude}}, \ and\ \bibinfo {author}
  {\bibfnamefont {M.}~\bibnamefont {Potemski}},\ }\href@noop {} {\bibfield
  {journal} {\bibinfo  {journal} {Phys. Rev. Lett.}\ }\textbf {\bibinfo
  {volume} {102}},\ \bibinfo {pages} {166401} (\bibinfo {year}
  {2009})}\BibitemShut {NoStop}%
\bibitem [{\citenamefont {Ando}(2007)}]{Ando2007}%
  \BibitemOpen
  \bibfield  {author} {\bibinfo {author} {\bibfnamefont {T.}~\bibnamefont
  {Ando}},\ }\href@noop {} {\bibfield  {journal} {\bibinfo  {journal} {J. Phys.
  Soc. Jpn.}\ }\textbf {\bibinfo {volume} {76}},\ \bibinfo {pages} {024712}
  (\bibinfo {year} {2007})}\BibitemShut {NoStop}%
\bibitem [{\citenamefont {Goerbig}\ \emph {et~al.}(2007)\citenamefont
  {Goerbig}, \citenamefont {Fuchs}, \citenamefont {Kechedzhi},\ and\
  \citenamefont {Fal'ko}}]{Goerbig2007}%
  \BibitemOpen
  \bibfield  {author} {\bibinfo {author} {\bibfnamefont {M.~O.}\ \bibnamefont
  {Goerbig}}, \bibinfo {author} {\bibfnamefont {J.-N.}\ \bibnamefont {Fuchs}},
  \bibinfo {author} {\bibfnamefont {K.}~\bibnamefont {Kechedzhi}}, \ and\
  \bibinfo {author} {\bibfnamefont {V.~I.}\ \bibnamefont {Fal'ko}},\
  }\href@noop {} {\bibfield  {journal} {\bibinfo  {journal} {Phys. Rev. Lett.}\
  }\textbf {\bibinfo {volume} {99}},\ \bibinfo {pages} {087402} (\bibinfo
  {year} {2007})}\BibitemShut {NoStop}%
\bibitem [{\citenamefont {Faugeras}\ \emph {et~al.}(2009)\citenamefont
  {Faugeras}, \citenamefont {Amado}, \citenamefont {Kossacki}, \citenamefont
  {Orlita}, \citenamefont {Sprinkle}, \citenamefont {Berger}, \citenamefont
  {de~Heer},\ and\ \citenamefont {Potemski}}]{Faugeras2009}%
  \BibitemOpen
  \bibfield  {author} {\bibinfo {author} {\bibfnamefont {C.}~\bibnamefont
  {Faugeras}}, \bibinfo {author} {\bibfnamefont {M.}~\bibnamefont {Amado}},
  \bibinfo {author} {\bibfnamefont {P.}~\bibnamefont {Kossacki}}, \bibinfo
  {author} {\bibfnamefont {M.}~\bibnamefont {Orlita}}, \bibinfo {author}
  {\bibfnamefont {M.}~\bibnamefont {Sprinkle}}, \bibinfo {author}
  {\bibfnamefont {C.}~\bibnamefont {Berger}}, \bibinfo {author} {\bibfnamefont
  {W.~A.}\ \bibnamefont {de~Heer}}, \ and\ \bibinfo {author} {\bibfnamefont
  {M.}~\bibnamefont {Potemski}},\ }\href {\doibase
  10.1103/PhysRevLett.103.186803} {\bibfield  {journal} {\bibinfo  {journal}
  {Phys. Rev. Lett.}\ }\textbf {\bibinfo {volume} {103}},\ \bibinfo {pages}
  {186803} (\bibinfo {year} {2009})}\BibitemShut {NoStop}%
\bibitem [{\citenamefont {Kim}\ \emph {et~al.}(2012)\citenamefont {Kim},
  \citenamefont {Ma}, \citenamefont {Imambekov}, \citenamefont {Kalugin},
  \citenamefont {Lombardo}, \citenamefont {Ferrari}, \citenamefont {Kono},\
  and\ \citenamefont {Smirnov}}]{Kim2012}%
  \BibitemOpen
  \bibfield  {author} {\bibinfo {author} {\bibfnamefont {Y.}~\bibnamefont
  {Kim}}, \bibinfo {author} {\bibfnamefont {Y.}~\bibnamefont {Ma}}, \bibinfo
  {author} {\bibfnamefont {A.}~\bibnamefont {Imambekov}}, \bibinfo {author}
  {\bibfnamefont {N.~G.}\ \bibnamefont {Kalugin}}, \bibinfo {author}
  {\bibfnamefont {A.}~\bibnamefont {Lombardo}}, \bibinfo {author}
  {\bibfnamefont {A.~C.}\ \bibnamefont {Ferrari}}, \bibinfo {author}
  {\bibfnamefont {J.}~\bibnamefont {Kono}}, \ and\ \bibinfo {author}
  {\bibfnamefont {D.}~\bibnamefont {Smirnov}},\ }\href {\doibase
  10.1103/PhysRevB.85.121403} {\bibfield  {journal} {\bibinfo  {journal} {Phys.
  Rev. B}\ }\textbf {\bibinfo {volume} {85}},\ \bibinfo {pages} {121403}
  (\bibinfo {year} {2012})}\BibitemShut {NoStop}%
\bibitem [{\citenamefont {Kashuba}\ and\ \citenamefont
  {Fal'ko}(2009)}]{Kashuba2009}%
  \BibitemOpen
  \bibfield  {author} {\bibinfo {author} {\bibfnamefont {O.}~\bibnamefont
  {Kashuba}}\ and\ \bibinfo {author} {\bibfnamefont {V.~I.}\ \bibnamefont
  {Fal'ko}},\ }\href {\doibase 10.1103/PhysRevB.80.241404} {\bibfield
  {journal} {\bibinfo  {journal} {Phys. Rev. B}\ }\textbf {\bibinfo {volume}
  {80}},\ \bibinfo {pages} {241404} (\bibinfo {year} {2009})}\BibitemShut
  {NoStop}%
\bibitem [{\citenamefont {Faugeras}\ \emph {et~al.}(2010)\citenamefont
  {Faugeras}, \citenamefont {Kossacki}, \citenamefont {Basko}, \citenamefont
  {Amado}, \citenamefont {Sprinkle}, \citenamefont {Berger}, \citenamefont
  {de~Heer},\ and\ \citenamefont {Potemski}}]{Faugeras2010a}%
  \BibitemOpen
  \bibfield  {author} {\bibinfo {author} {\bibfnamefont {C.}~\bibnamefont
  {Faugeras}}, \bibinfo {author} {\bibfnamefont {P.}~\bibnamefont {Kossacki}},
  \bibinfo {author} {\bibfnamefont {D.~M.}\ \bibnamefont {Basko}}, \bibinfo
  {author} {\bibfnamefont {M.}~\bibnamefont {Amado}}, \bibinfo {author}
  {\bibfnamefont {M.}~\bibnamefont {Sprinkle}}, \bibinfo {author}
  {\bibfnamefont {C.}~\bibnamefont {Berger}}, \bibinfo {author} {\bibfnamefont
  {W.~A.}\ \bibnamefont {de~Heer}}, \ and\ \bibinfo {author} {\bibfnamefont
  {M.}~\bibnamefont {Potemski}},\ }\href {\doibase 10.1103/PhysRevB.81.155436}
  {\bibfield  {journal} {\bibinfo  {journal} {Phys. Rev. B}\ }\textbf {\bibinfo
  {volume} {81}},\ \bibinfo {pages} {155436} (\bibinfo {year}
  {2010})}\BibitemShut {NoStop}%
\bibitem [{\citenamefont {Ferrari}\ \emph {et~al.}(2006)\citenamefont
  {Ferrari}, \citenamefont {Meyer}, \citenamefont {Scardaci}, \citenamefont
  {Casiraghi}, \citenamefont {Lazzeri}, \citenamefont {Mauri}, \citenamefont
  {Piscanec}, \citenamefont {Jiang}, \citenamefont {Novoselov}, \citenamefont
  {Roth},\ and\ \citenamefont {Geim}}]{Ferrari2006}%
  \BibitemOpen
  \bibfield  {author} {\bibinfo {author} {\bibfnamefont {A.~C.}\ \bibnamefont
  {Ferrari}}, \bibinfo {author} {\bibfnamefont {J.~C.}\ \bibnamefont {Meyer}},
  \bibinfo {author} {\bibfnamefont {V.}~\bibnamefont {Scardaci}}, \bibinfo
  {author} {\bibfnamefont {C.}~\bibnamefont {Casiraghi}}, \bibinfo {author}
  {\bibfnamefont {M.}~\bibnamefont {Lazzeri}}, \bibinfo {author} {\bibfnamefont
  {F.}~\bibnamefont {Mauri}}, \bibinfo {author} {\bibfnamefont
  {S.}~\bibnamefont {Piscanec}}, \bibinfo {author} {\bibfnamefont
  {D.}~\bibnamefont {Jiang}}, \bibinfo {author} {\bibfnamefont {K.~S.}\
  \bibnamefont {Novoselov}}, \bibinfo {author} {\bibfnamefont {S.}~\bibnamefont
  {Roth}}, \ and\ \bibinfo {author} {\bibfnamefont {A.~K.}\ \bibnamefont
  {Geim}},\ }\href@noop {} {\bibfield  {journal} {\bibinfo  {journal} {Phys.
  Rev. Lett.}\ }\textbf {\bibinfo {volume} {97}},\ \bibinfo {pages} {187401}
  (\bibinfo {year} {2006})}\BibitemShut {NoStop}%
\bibitem [{\citenamefont {Ni}\ \emph {et~al.}(2008)\citenamefont {Ni},
  \citenamefont {Wang}, \citenamefont {Yu}, \citenamefont {You},\ and\
  \citenamefont {Shen}}]{Ni2008}%
  \BibitemOpen
  \bibfield  {author} {\bibinfo {author} {\bibfnamefont {Z.}~\bibnamefont
  {Ni}}, \bibinfo {author} {\bibfnamefont {Y.}~\bibnamefont {Wang}}, \bibinfo
  {author} {\bibfnamefont {T.}~\bibnamefont {Yu}}, \bibinfo {author}
  {\bibfnamefont {Y.}~\bibnamefont {You}}, \ and\ \bibinfo {author}
  {\bibfnamefont {Z.}~\bibnamefont {Shen}},\ }\href@noop {} {\bibfield
  {journal} {\bibinfo  {journal} {Phys. Rev. B}\ }\textbf {\bibinfo {volume}
  {77}},\ \bibinfo {pages} {235403} (\bibinfo {year} {2008})}\BibitemShut
  {NoStop}%
\bibitem [{\citenamefont {Carozo}\ \emph {et~al.}(2011)\citenamefont {Carozo},
  \citenamefont {Almeidad}, \citenamefont {Ferreira}, \citenamefont {Cancado},
  \citenamefont {Achete},\ and\ \citenamefont {Jorio}}]{Carozo2011}%
  \BibitemOpen
  \bibfield  {author} {\bibinfo {author} {\bibfnamefont {V.}~\bibnamefont
  {Carozo}}, \bibinfo {author} {\bibfnamefont {C.}~\bibnamefont {Almeidad}},
  \bibinfo {author} {\bibfnamefont {E.}~\bibnamefont {Ferreira}}, \bibinfo
  {author} {\bibfnamefont {L.}~\bibnamefont {Cancado}}, \bibinfo {author}
  {\bibfnamefont {C.}~\bibnamefont {Achete}}, \ and\ \bibinfo {author}
  {\bibfnamefont {A.}~\bibnamefont {Jorio}},\ }\href@noop {} {\bibfield
  {journal} {\bibinfo  {journal} {NanoLett.}\ }\textbf {\bibinfo {volume}
  {11}},\ \bibinfo {pages} {4527} (\bibinfo {year} {2011})}\BibitemShut
  {NoStop}%
\bibitem [{\citenamefont {He}\ \emph {et~al.}(2013)\citenamefont {He},
  \citenamefont {Chung}, \citenamefont {Delaney}, \citenamefont {Keiser},
  \citenamefont {Jauregui}, \citenamefont {Shand}, \citenamefont {Chancey},
  \citenamefont {Wang}, \citenamefont {Bao},\ and\ \citenamefont
  {Chen}}]{He2013}%
  \BibitemOpen
  \bibfield  {author} {\bibinfo {author} {\bibfnamefont {R.}~\bibnamefont
  {He}}, \bibinfo {author} {\bibfnamefont {T.-C.}\ \bibnamefont {Chung}},
  \bibinfo {author} {\bibfnamefont {C.}~\bibnamefont {Delaney}}, \bibinfo
  {author} {\bibfnamefont {C.}~\bibnamefont {Keiser}}, \bibinfo {author}
  {\bibfnamefont {L.~A.}\ \bibnamefont {Jauregui}}, \bibinfo {author}
  {\bibfnamefont {P.~M.}\ \bibnamefont {Shand}}, \bibinfo {author}
  {\bibfnamefont {C.~C.}\ \bibnamefont {Chancey}}, \bibinfo {author}
  {\bibfnamefont {Y.}~\bibnamefont {Wang}}, \bibinfo {author} {\bibfnamefont
  {J.}~\bibnamefont {Bao}}, \ and\ \bibinfo {author} {\bibfnamefont {Y.~P.}\
  \bibnamefont {Chen}},\ }\href@noop {} {\bibfield  {journal} {\bibinfo
  {journal} {NanoLett.}\ }\textbf {\bibinfo {volume} {13}},\ \bibinfo {pages}
  {3594} (\bibinfo {year} {2013})}\BibitemShut {NoStop}%
\bibitem [{\citenamefont {Hwang}\ \emph {et~al.}(2012)\citenamefont {Hwang},
  \citenamefont {Siegel}, \citenamefont {Mo}, \citenamefont {Regan},
  \citenamefont {Ismach}, \citenamefont {Zhang}, \citenamefont {A.},\ and\
  \citenamefont {Lanzara}}]{Hwang2012}%
  \BibitemOpen
  \bibfield  {author} {\bibinfo {author} {\bibfnamefont {C.}~\bibnamefont
  {Hwang}}, \bibinfo {author} {\bibfnamefont {D.~A.}\ \bibnamefont {Siegel}},
  \bibinfo {author} {\bibfnamefont {S.~K.}\ \bibnamefont {Mo}}, \bibinfo
  {author} {\bibfnamefont {W.}~\bibnamefont {Regan}}, \bibinfo {author}
  {\bibfnamefont {A.}~\bibnamefont {Ismach}}, \bibinfo {author} {\bibfnamefont
  {Y.}~\bibnamefont {Zhang}}, \bibinfo {author} {\bibfnamefont {A.~Z.}\
  \bibnamefont {A.}}, \ and\ \bibinfo {author} {\bibnamefont {Lanzara}},\
  }\href@noop {} {\bibfield  {journal} {\bibinfo  {journal} {Sci. Rep.}\
  }\textbf {\bibinfo {volume} {2}},\ \bibinfo {pages} {590} (\bibinfo {year}
  {2012})}\BibitemShut {NoStop}%
\bibitem [{\citenamefont {dos Santos}\ \emph {et~al.}(2007)\citenamefont {dos
  Santos}, \citenamefont {Peres},\ and\ \citenamefont {Neto}}]{Lopes07}%
  \BibitemOpen
  \bibfield  {author} {\bibinfo {author} {\bibfnamefont {J.~L.}\ \bibnamefont
  {dos Santos}}, \bibinfo {author} {\bibfnamefont {N.~M.~R.}\ \bibnamefont
  {Peres}}, \ and\ \bibinfo {author} {\bibfnamefont {A.~H.~C.}\ \bibnamefont
  {Neto}},\ }\href@noop {} {\bibfield  {journal} {\bibinfo  {journal} {Phys.
  Rev. Lett.}\ }\textbf {\bibinfo {volume} {99}},\ \bibinfo {pages} {256802}
  (\bibinfo {year} {2007})}\BibitemShut {NoStop}%
\bibitem [{\citenamefont {de~Laissardiere}\ \emph {et~al.}(2010)\citenamefont
  {de~Laissardiere}, \citenamefont {Mayou},\ and\ \citenamefont
  {Magaud}}]{Trambly2010}%
  \BibitemOpen
  \bibfield  {author} {\bibinfo {author} {\bibfnamefont {G.~T.}\ \bibnamefont
  {de~Laissardiere}}, \bibinfo {author} {\bibfnamefont {D.}~\bibnamefont
  {Mayou}}, \ and\ \bibinfo {author} {\bibfnamefont {L.}~\bibnamefont
  {Magaud}},\ }\href@noop {} {\bibfield  {journal} {\bibinfo  {journal}
  {NanoLett.}\ }\textbf {\bibinfo {volume} {10}},\ \bibinfo {pages} {804}
  (\bibinfo {year} {2010})}\BibitemShut {NoStop}%
\bibitem [{\citenamefont {Can\ifmmode~\mbox{\c{c}}\else \c{c}\fi{}ado}\ \emph
  {et~al.}(2007)\citenamefont {Can\ifmmode~\mbox{\c{c}}\else \c{c}\fi{}ado},
  \citenamefont {Jorio},\ and\ \citenamefont {Pimenta}}]{Pimenta2007}%
  \BibitemOpen
  \bibfield  {author} {\bibinfo {author} {\bibfnamefont {L.~G.}\ \bibnamefont
  {Can\ifmmode~\mbox{\c{c}}\else \c{c}\fi{}ado}}, \bibinfo {author}
  {\bibfnamefont {A.}~\bibnamefont {Jorio}}, \ and\ \bibinfo {author}
  {\bibfnamefont {M.~A.}\ \bibnamefont {Pimenta}},\ }\href {\doibase
  10.1103/PhysRevB.76.064304} {\bibfield  {journal} {\bibinfo  {journal} {Phys.
  Rev. B}\ }\textbf {\bibinfo {volume} {76}},\ \bibinfo {pages} {064304}
  (\bibinfo {year} {2007})}\BibitemShut {NoStop}%
\bibitem [{\citenamefont {Basko}(2009)}]{Basko2009}%
  \BibitemOpen
  \bibfield  {author} {\bibinfo {author} {\bibfnamefont {D.~M.}\ \bibnamefont
  {Basko}},\ }\href@noop {} {\bibfield  {journal} {\bibinfo  {journal} {New. J.
  Phys.}\ }\textbf {\bibinfo {volume} {11}},\ \bibinfo {pages} {095011}
  (\bibinfo {year} {2009})}\BibitemShut {NoStop}%
\bibitem [{\citenamefont {Cai}\ \emph {et~al.}(2010)\citenamefont {Cai},
  \citenamefont {Moore}, \citenamefont {Zhu}, \citenamefont {Li}, \citenamefont
  {Chen}, \citenamefont {Shi},\ and\ \citenamefont {Ruoff}}]{Cai2010}%
  \BibitemOpen
  \bibfield  {author} {\bibinfo {author} {\bibfnamefont {W.}~\bibnamefont
  {Cai}}, \bibinfo {author} {\bibfnamefont {A.~L.}\ \bibnamefont {Moore}},
  \bibinfo {author} {\bibfnamefont {Y.}~\bibnamefont {Zhu}}, \bibinfo {author}
  {\bibfnamefont {X.}~\bibnamefont {Li}}, \bibinfo {author} {\bibfnamefont
  {S.}~\bibnamefont {Chen}}, \bibinfo {author} {\bibfnamefont {L.}~\bibnamefont
  {Shi}}, \ and\ \bibinfo {author} {\bibfnamefont {R.~S.}\ \bibnamefont
  {Ruoff}},\ }\href@noop {} {\bibfield  {journal} {\bibinfo  {journal}
  {NanoLett.}\ }\textbf {\bibinfo {volume} {10}},\ \bibinfo {pages} {1645}
  (\bibinfo {year} {2010})}\BibitemShut {NoStop}%
\end{thebibliography}
\end{document}